\documentclass[traditabstract]{aa} 
\usepackage{graphicx}
\usepackage{txfonts}
\usepackage{natbib}

\def\yr{{\rm yr}}
\def\second{{\rm sec}}
\def\km{{\rm km}}
\def\Mpc{{\rm Mpc}}

\begin{document}

\title{The Core-collapse rate from  the Supernova Legacy Survey}

\author{G. Bazin\inst{1}, N. Palanque-Delabrouille\inst{1},J. Rich\inst{1},  V. Ruhlmann-Kleider\inst{1}, 
E. Aubourg\inst{1,2}, L. Le Guillou\inst{3},
P. Astier\inst{3}, C. Balland\inst{3,4},
S. Basa\inst{5},
 R. G. Carlberg\inst{7}, A. Conley\inst{7}, D. Fouchez\inst{8}, J. Guy\inst{3},
D. Hardin\inst{3}, I. M. Hook\inst{10}, D. A. Howell\inst{7},
R. Pain\inst{3}, K. Perrett\inst{7},
C. J. Pritchet\inst{9}, N.  Regnault\inst{3}, 
M. Sullivan\inst{10}, P. Antilogus\inst{3},
V. Arsenijevic\inst{11}, S. Baumont\inst{3}, S. Fabbro\inst{11},
J. Le Du\inst{8}, C. Lidman\inst{12}, M. Mouchet\inst{2,6},
A. Mour\~ao\inst{11}, E. S. Walker\inst{10}
}

\institute{ 
CEA/Saclay, DSM/Irfu/Spp, F-91191 Gif-sur-Yvette Cedex, France
\and APC, UMR 7164 CNRS, 10 rue Alice Domon et L\'eonie Duquet, F-75205 Paris Cedex 13, France
\and LPNHE, CNRS-IN2P3 and Universities of Paris 6 \& 7,F-75252 Paris Cedex 05, France
\and University Paris 11, Orsay, F-91405, France
\and LAM, CNRS, BP8, P\^{o}le de l'\'etoile, Site de Ch\^{a}teau-Gombert,
38 rue Fr\'ed\'eric Joliot-Curie, F-13388 Marseille Cedex 13, France
\and LUTH, UMR 8102 CNRS, Observatoire de Paris, Section de Meudon, F-92195 Meudon Cedex, France
\and Deparment of Astronomy and Astrophysics, University of Toronto, 50 St. George Street, Toronto, ON M5S 3H8, Canada
\and CPPM, CNRS-Luminy, Case 907, F-13288 Marseille Cedex 9, France
\and Department of Physics and Astronomy, University of Victoria, PO Box 3055, Victoria, BC V8W 3P6, Canada
\and University of Oxford, Astrophysics, Denys Wilkinson Building,Kneble Road, Oxford OX1 3RH, UK
\and CENTRA-Centro M. de Astrofisica and Department of Physics, IST, Lisbon, Portugal
\and European Southern Observatory, Alonso de Cordova 3107, Vitacura, Casilla 19001, Santiago 19, Chile
}

   \date{Received June 15, 2008; accepted yyyy xx, 2008}


\authorrunning{G. Bazin et al.}
\titlerunning{The Core-collapse Rate from 
 SNLS}
 
  \abstract{
We use three years of data from the Supernova Legacy Survey (SNLS)
to study the general properties of core-collapse  and type Ia supernovae.
This is the first such study using the ``rolling search'' technique
which guarantees well-sampled SNLS light curves and good efficiency for
supernovae brighter than $i^\prime\sim24$.
Using host photometric redshifts,
we measure the supernova absolute magnitude distribution 
down to luminosities $4.5\,{\rm mag}$ fainter than normal SNIa.
Using spectroscopy and light-curve fitting to
discriminate against SNIa,
we find a sample of 117 core-collapse supernova candidates with
redshifts $z<0.4$  
(median redshift of 0.29)
and measure their rate  to be  larger than
the type Ia supernova rate by a factor $4.5\pm0.8(stat.)\, \pm0.6 (sys.)$.
This corresponds to a core-collapse rate at $z=0.3$ of 
$[1.42\pm 0.3(stat.)\,\pm0.3(sys.)]\times10^{-4}\yr^{-1}(h_{70}^{-1}\Mpc)^{-3}$.
}
  

   \maketitle
%

\section{Introduction}

The rate of supernova explosions is astrophysically
important because it
determines the rate at which
heavy elements are dispersed into the interstellar medium,
thereby constraining galactic chemical evolution.
Since the progenitors of core-collapse supernovae (SNcc) are 
believed to be short-lived massive stars,
the SNcc rate is expected  to reflect the star-formation
rate, increasing with redshift
like  $\sim(1+z)^{3.6}$ for $z<0.5$  \citep{hopkins}.
Thermonuclear type Ia supernovae
(SNIa) have both long- and short-lived progenitors so the
SNIa rate has a delayed component making the  
SNIa rate rise more slowly with redshift, 
$\sim(1+z)^{2}$ \citep{pritchet}.

The SNIa rate is now known to a precision of about $20\%$.
Measurements have profited from  
the  high luminosity of SNIa which 
make them  relatively easy to detect and identify.  
Furthermore,
their utility as cosmological  distance indicators has
motivated intense searches.  
An example is  the Supernova Legacy
Survey (SNLS) at the Canada-France-Hawaii Telescope (CFHT)
performed between 2003 and 2008.
Using early SNLS data, \citet{neill2006}
derived   a SNIa rate at a redshift $z\sim 0.5$ of
\begin{displaymath}
\frac{R_{Ia}(z=0.5)}
{10^{-4} \yr^{-1}(h_{70}^{-1}\Mpc)^{-3}}
\,=\,
0.42\pm 0.06(stat.)\,^{+0.13}_{-0.09}(syst.)
\end{displaymath}
where $h_{70}=H_0/70\km\,\second^{-1}\Mpc^{-1}$.

The rate for SNcc is more 
difficult to measure because observed SNcc have a magnitude
distribution that peaks roughly
$1.5\,{\rm mag}$ fainter than SNIa
and covers a range of more than $5\,{\rm mag}$ \citep{richardson}.
The local rate was measured by \citet{Capp1999}
using  137 supernovae discovered by eye and photographically.
Most had spectroscopic identification, about half being SNIa 
and half SNcc  (SNIb/c and SNII).
After efficiency corrections, the
SNcc rate was found to be a factor  $2.4\pm1.3$ greater
than the SNIa rate. 

The SNcc rate at $z\sim0.3$ was measured by \citet{Capp2005} and more
recently by \citet{Bott2008}.
The latter  used images taken over a six year period with typically
four months between images.
They found 18 SNcc candidates 
and 13 SNIa candidates 
(of which a total of 25 are spectroscopically confirmed)
to find a SNcc rate at $z\sim0.26$ a  factor $4\pm2$ greater than
the SNIa rate.
Finally, \citet{dahlen} used the
Advanced Camera for Surveys 
on the Hubble Space Telescope to obtain
images for five epochs separated by $\sim45\,{\rm days}$.
For redshifts $<1$, they found
17 SNIa candidates (with some spectroscopic identification) 
and 16 SNcc candidates (no spectroscopic identification) 
which allowed them to derive
$R_{cc}/R_{Ia}=3.6\pm2.0$ at $z\sim0.4$ and
$R_{cc}/R_{Ia}=2.5\pm1.0$ at $z\sim0.8$.
 
All existing measurements of the SNcc rate suffer from the
fact that the discovery procedure involved the comparison
of images separated in time by intervals comparable to
or greater than the characteristic $\sim 1\,{\rm month}$
time scales of supernovae.
Consequently, 
well-sampled
light curves for most 
candidates are not available,
complicating the type identification  and
efficiency calculations.
The SNLS ``rolling search''   avoids this problem
because of its high cadence 
monitoring of four $1\,deg^2$ fields in the $g^\prime$, 
$r^\prime$, $i^\prime$ and $z^\prime$ bands
over a total of 5 years.
During each 6 month observing season for each field,
typically four observations per lunation were obtained in the 
$r^\prime$ and $i^\prime$ bands,
three in the $z^\prime$ band  and two in the $g^\prime$ band.
This strategy yields well-sampled light curves 
(e.g. Figs. \ref{lcIa}, \ref{lcplateau} and
\ref{lcCC}) with high efficiency for all events 
occurring during the observing season and having maximum fluxes
brighter than $i^\prime\sim24$.
This makes the sample of normal SNIa essentially complete up to $z=0.6$.
For the fainter SNcc, SNLS effectively monitors a volume that
is a calculable function of the apparent magnitude and redshift.
This will allow us to derive 
the differential supernova rate 
(rate per absolute magnitude interval) for
supernova with redshifts $<0.4$ and 
within $4.5\,{\rm mag}$ of normal SNIa.

The primary goal of SNLS was cosmology with SNIa.
As such, mostly  SNIa-like objects were
targeted for spectroscopy \citep{sullivan2006} and
the majority of our SN candidates do not have
spectroscopic identification or redshifts.
We therefore used host photometric redshifts for this
study though we are in the process of obtaining
host spectroscopic redshifts.
For supernovae without spectroscopic identification,
knowledge of the host redshift allows us to determine
if the supernova four-band light-curves are
consistent with the family of light curves typical of
SNIa.
The combination of spectroscopic and photometric typing
will allow us to 
identify most SNIa.
A relatively uncontaminated sample of SNcc 
is then defined as those supernovae
not identified as SNIa.
Use of the previously measured SNIa rate \citep{neill2006} 
will then allow us
to derive the SNcc rate.
The measurement will use only supernovae with redshifts
$<0.4$, beyond which the efficiency for detecting SNcc
is too small to add significantly to the sample.
This has the additional advantage that in this redshift 
range, the 615nm Si II absorption feature is visible
simplifying spectroscopic identification of SNIa.

The outline of this paper is as follows.  Section \ref{evselecsec} presents
the light curve construction and event selection.
Section \ref{charactsec} presents the characteristics of the supernova
candidates.
Section \ref{classificationsec} defines the SNIa and SNcc candidate
samples.
Section \ref{ratesec} derives the relative SNIa and SNcc rates
from which we deduce the SNcc rate.
Section \ref{sumsec} concludes with a comparison of previous results.

Throughout, magnitudes are expressed in the AB system 
\citep{abmagref}.
A flat $\Lambda CDM$ universe with $\Omega_M=0.27$ is assumed.

\section{Event selection}\label{evselecsec}

For this study, we performed a ``deferred'' search for transient events
that was completely independent of the 
real-time search\footnote{http://legacy.astro.utoronto.ca/} used
to select supernovae for spectroscopy 
targets and for subsequent use in cosmological
parameter analyzes.
The details of the deferred search are given elsewhere 
\citep{bib:gbMC,gbinprep}.
We used
SNLS observations of the four ``deep'' fields (D1,D2,D3,D4)
from January 1st, 2003 to September 21, 2006. 
A reference image for each field and filter was constructed by co-adding
the images from 20 good quality nights.
The reference image was then subtracted from all science images of the same 
field and filter (after seeing-adjustment).
In the  $i^\prime$ filter,
the subtracted images from each lunation were combined to form one
``stacked'' image 
per lunation
and stellar objects were searched for on each of these stacks.
Approximately 300,000 objects were found, 
mostly spurious detections due to saturated signals from bright objects.
Four-filter light curves 
for these objects were then obtained from individual subtracted images by
differential photometry with PSF fitting, imposing the position
found on the $i^\prime$ stack.
Fluxes were calibrated using the set of SNLS tertiary
standards \citep{bib:astier}.

The event selection criteria applied on the detected light curves are
described in detail in~\citet{gbinprep}.
Spurious detections were mostly eliminated by requiring that the
light curves in $i^\prime$ and $r^\prime$ have at least
three successive photometric points with fluxes above $1\sigma$ from
base line and their dates of maximum flux should be within 50 days from 
each other. 
Light curves corresponding to detections near stars as identified 
in our reference images were also discarded. 
Accepted light curves were  fit with the phenomenological form
\begin{equation}
f(t)\;=\; A \frac{e^{-(t-t_0)/\tau_{fall}}}{1+e^{(t-t_0)/\tau_{rise}}} \;+ B
\label{fitformula}
\end{equation}
While this form has no particular physical motivation, it is
sufficiently general to fit the shape of all types of supernovae.
Long-term variable objects
(such as AGNs) were rejected by comparing the $\chi^2$ of the light
curve fit using (\ref{fitformula}) 
 with fits to a constant flux, and only accepting objects
for which the phenomenological model is a substantially better fit.
In addition, we require that the light-curve be consistent with
a time-independent flux before and after the main variation as fitted
by (\ref{fitformula}).
The precise cuts were defined with the help of synthetic 
SNIa light curves and selected real light curves which have been
confirmed by spectroscopy as type Ia or core-collapse SNe.
Finally, good time sampling criteria were applied, i.e. requiring
at least 
one pre-max epoch within 30 days
and one post-max epoch within 60 days of   the date of
maximum flux in the $i^\prime$ and $r^\prime$ filters, and
at least two epochs 
in that time interval in the $g^\prime$ and $z^\prime$ filters.
A set of 1462 events was thus retained.

\begin{figure}
\includegraphics[width=8.5cm]{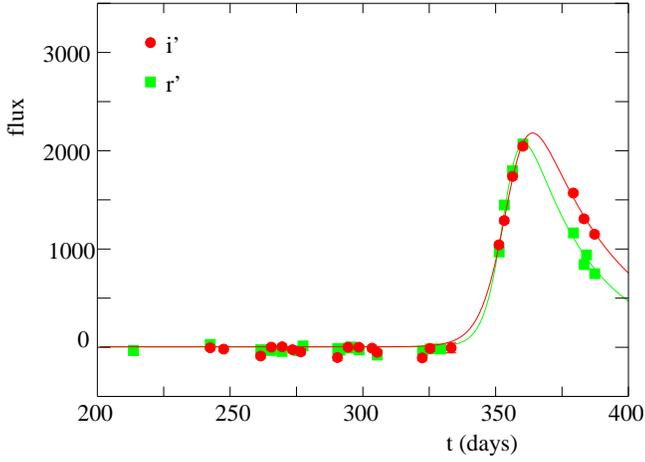}
\caption{ 
The $r^\prime$ and $i^\prime$ 
light curves of a SNIa. 
The spectroscopic redshift is 0.332,
the host photometric redshift 0.294, and  the peak
magnitude  $i^\prime=21.6$.
The time corresponds to MJD-52640.
}
\label{lcIa}
\end{figure}

\begin{figure}
\includegraphics[width=8.5cm]{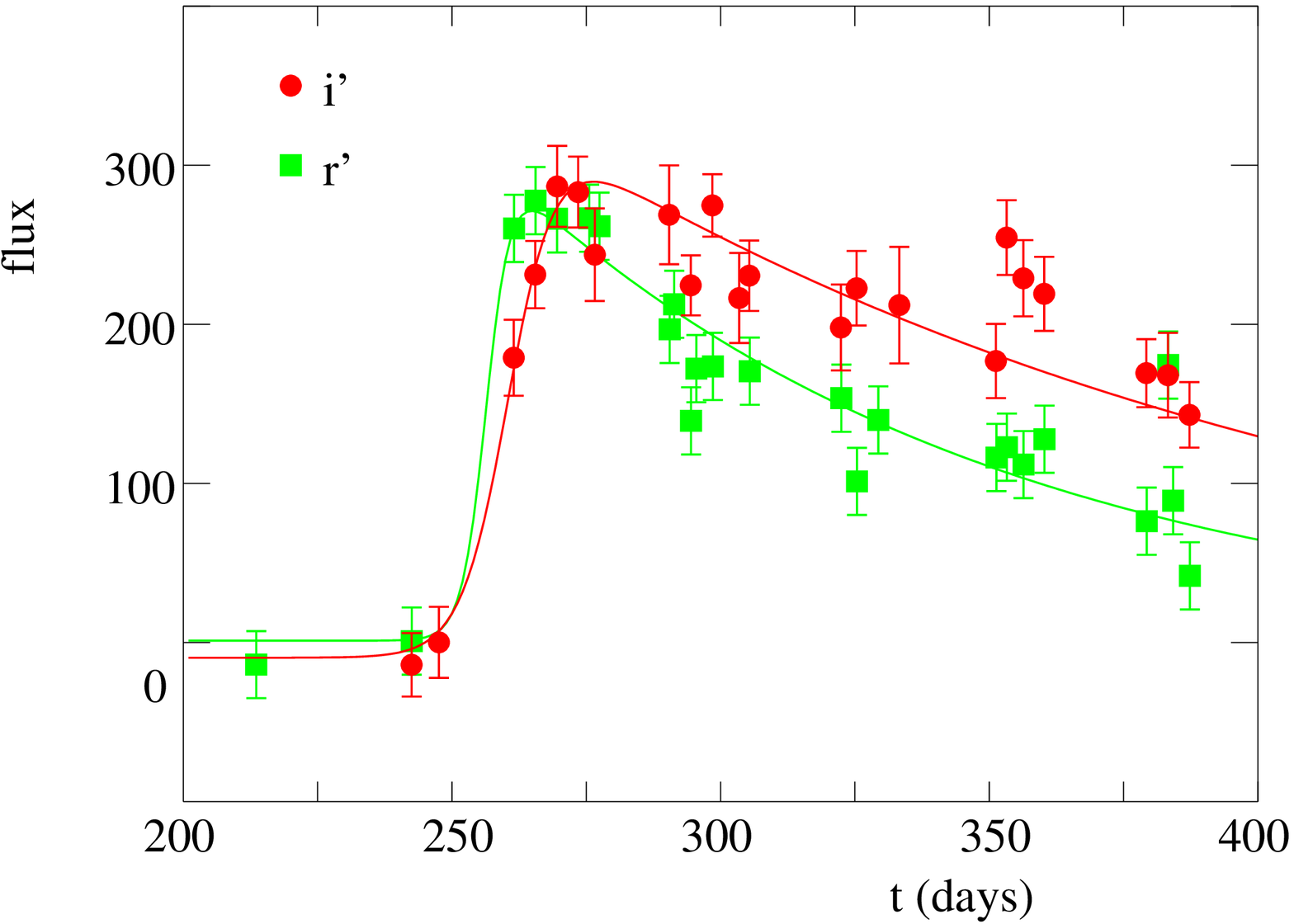}
\caption{ 
The $r^\prime$ and $i^\prime$ 
light curves of a SNIIp. The spectroscopic redshift is 0.328,       
the host photometric redshift  0.335, and the peak magnitude $i^\prime=23.8$.
The time corresponds to MJD-52640.
}
\label{lcplateau}
\end{figure}

\begin{figure}
\includegraphics[width=8.5cm]{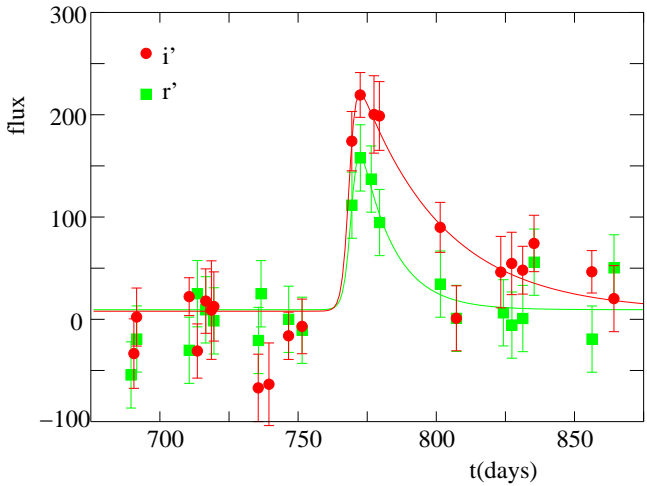}
\caption{ 
The  $r^\prime$ and $i^\prime$ 
light curves of a SNcc candidate.  The host photometric
redshift is 0.36 and the peak magniude  $i^\prime=24.1$.  
It is the faintest event used here to measure the core-collapse rate.
The time corresponds to MJD-52640.
}
\label{lcCC}
\end{figure}

Light-curves for three  events are shown in 
Figs. \ref{lcIa},\ref{lcplateau} and \ref{lcCC}.
The first shows a typical spectroscopically-confirmed SNIa
with spectroscopic redshift $z=0.332$
and the second a typical spectroscopically-confirmed SNcc
with spectroscopic redshift $z=0.328$.
The third is one of the faintest events to be used in Section \ref{ratesec}
to measure the core-collapse rate.
Its peak magnitude is $i^\prime=24.1$, as fitted by~\ref{fitformula}. 

\begin{figure}
\includegraphics[width=8.5cm]{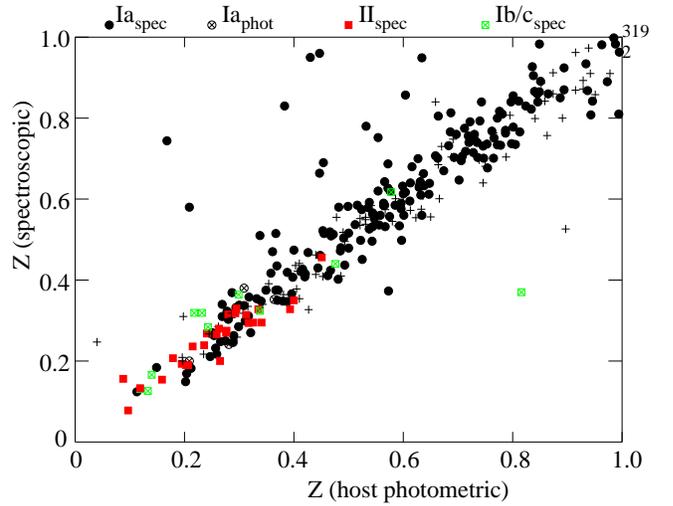}
\caption{ 
SNLS spectroscopic redshifts vs. host photometric redshifts taken
from the catalog of \citet{ilbert2006}.
Spectroscopically identified supernovae  and photometrically
identified SNIa are marked by the signs defined at the top of the figure.
The crosses are supernovae whose type is not determined
spectroscopically or
photometrically.  
}
\label{zvszfig}
\end{figure}

To identify host galaxies for the events, we used
the photometric galaxy catalog of \citet{ilbert2006}.
The host for an event was chosen to be the 
galaxy with the smallest distance, $r$, between the event and
the galaxy center in units of the galaxy's effective radius, $r_{\rm gal}$,
defined as
the half-width of the galaxy in the direction of the
event.
The value of $r_{\rm gal}$ was defined by the $A$, $B$ and $\theta$ SExtractor
parameters \citep{sextractor}.
The match was considered successful if the host was at a distance
$r<5r_{\rm gal}$.
This choice was a compromise between host finding efficiency
and accidental mismatching.
Of the 1462 selected events, 1329 (91\%) have matched hosts
and of these 1207 (91\%) have a photometric redshift.
Figure \ref{zvszfig} shows the host spectroscopic redshift vs.
photometric redshift \citep{ilbert2006} for events with both.
After elimination of outliers,
the deduced resolution for photometric redshifts is  $\sigma_z\sim0.04$
for $z<0.4$.

For the SNcc rate measurements, we will consider only the 
239
events with $z_{host}<0.4$.
These light curves were visually scanned in order to eliminate a few
residual non-SN events. Six light curves were clear physical variable
events, varying on a time-scale consistent with that of SNe but their
light curves showed other details incompatible with that hypothesis (no
flux in g',r',z' filter or rise time longer than fall time). Another 12
events had light curves with very low maximum flux and erratic
variations and thus most probably residual noise events which appear to
be associated with  low redshift galaxies.
After elimination of these events, we were left with 221 events. 

\begin{figure}
\includegraphics[width=8.5cm]{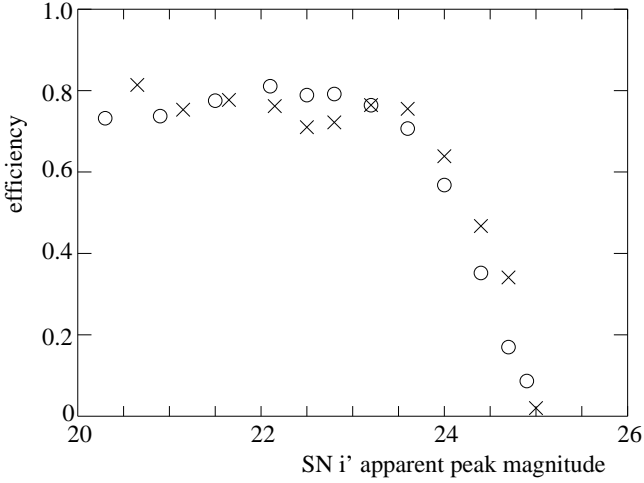}
\caption{ 
Pre-selection efficiency as a function of $i^\prime$.
The ``efficiency'' is defined as the number of reconstructed
events in a given magnitude interval divided by the number
of generated events in the same interval.
The circles are for SNIa and the crosses for SNcc with
$\tau_{fall}=100{\rm days}$.
}
\label{efffig}
\end{figure}

The efficiency of the event selection procedure was calculated 
by treating
simulated supernovae with the same procedure.
Supernovae added to real $i^\prime$-band images were used to
test the initial detection stage in $i^\prime$.
The efficiency of the subsequent event selection cuts was calculated
by applying them to  light curves
generated by a Monte-Carlo simulation that takes into
account the photometric resolution and the observing sequence.
The resulting efficiency is a function of the maximum
fluxes in the four bands and the associated time scales.
However, to good approximation the efficiency is
simply a function of the maximum in the $i^\prime$ band.
The efficiency is shown in Figure \ref{efffig} for
SNIa and for long SNcc ($\tau_{fall}=100\,{\rm days}$).
In both cases,
the efficiency is relatively $i^\prime$-independent 
at a value of $\sim0.8$ for
$i^\prime<23$ at which point it starts to decline, reaching
0.4 at $i^\prime=24.3$.

The performance of our selection pipeline was checked by comparing
it with the results of
the SNLS real-time pipeline used to select spectroscopy
targets.  A total of 340 supernovae 
were targeted during the period considered here
including events as faint as $i^\prime=24.4$.
Of these, all but two were found on the $i^\prime$
stacked images.  
(The two lost events were outside the  reference images.)
Of the 338 events, 295 passed our selection criteria.
The loss of the 43 events was due to our 
time sampling criteria which is more restrictive than
the real-time criteria.

\section{Event characteristics}
\label{charactsec}

\begin{figure}
\includegraphics[width=8.5cm]{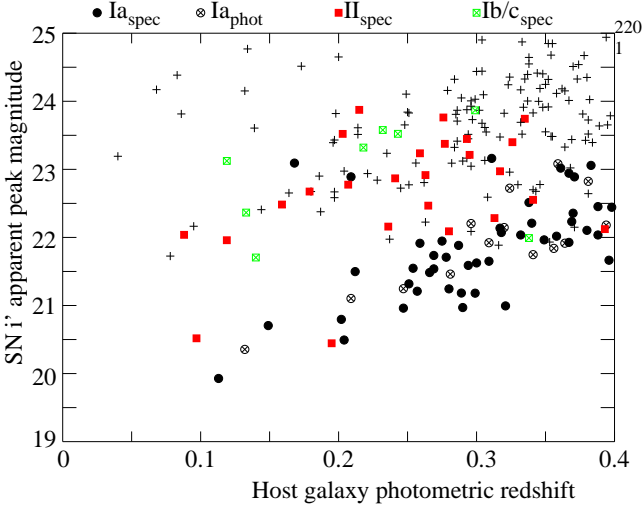}
\caption{ 
The peak $i^\prime$ magnitude vs. host photometric redshift.  
Spectroscopically identified supernovae  and photometrically
identified SNIa are marked by the signs defined at the top of the figure.
Events without such identification (crosses) are SNcc candidates
without spectroscopic confirmation.
Most identified SNIa lie in the
band with $i^\prime \sim 21.8+5\log(z/0.3)$.
Note that the two anomalously faint
spectroscopically confirmed SNIa 
at $z\sim0.17$ and $z\sim0.21$
are the corresponding outliers in Fig. \ref{zvszfig} indicating
an incorrect host-photometric redshift.
}
\label{hubbleifig}
\end{figure}

Figure \ref{hubbleifig}
shows the $i^\prime$  Hubble diagram for the 221 events
with host photometric redshifts $<0.4$.
Events that are spectroscopically
identified as SNIa or SNcc (SNII, SNIb, SNIc) are marked. 
Also marked are photometrically identified SNIa as
discussed in Sec. \ref{classificationsec}.
The spectroscopic SNIa's fall mostly 
along the band of bright events
centered approximately on $i^\prime\sim21.8+5\log(z/0.3)$.
The spectroscopic SNcc's are generally fainter with
$i^\prime>22.7+5\log(z/0.3)$.

The supernovae that we will use to measure rates have
a wide range of redshifts up to $z=0.4$. 
In order to compare supernovae of differing $z$, we
define an   AB magnitude centered on $570\,nm$
in the supernova rest-frame by a simple
redshift-dependent interpolation between $r^\prime$ and $i^\prime$:
\begin{equation}
m_{570} \;\equiv\; (4z-0.4)i^\prime \,+\, (1.4-4z)r^\prime \;.
\label{m570def}
\end{equation}
This gives 
$m_{570}(z=0.1)=r^\prime$  ($\lambda=626\,nm$) and
$m_{570}(z=0.35)=i^\prime$  ($\lambda=769\,nm$).
We then define the quantity $\Delta M_{570}$
to be proportional to the absolute magnitude
taking into 
account the supernova distance but not host absorption:
\begin{equation}
\Delta M_{570} \;\equiv\; m_{570} - 2.5\log[(1+z)d(z)^2] \,-\,C 
\label{M600eq}
\end{equation}
where
\begin{equation}
d(z) \;=\; \int_0^z 
\frac{dz^\prime }{\sqrt{\Omega_m (1+z^\prime )^3 + (1-\Omega_m)}}
\;=\frac{d_L}{(c/H_0)(1+z)} \;.
\label{ddefeq}
\end{equation}
The constant $C=24.2$ is chosen so that the spectroscopically
confirmed SNIa are centered on $\Delta M_{570}=0$.

\begin{figure}
\includegraphics[width=8.5cm]{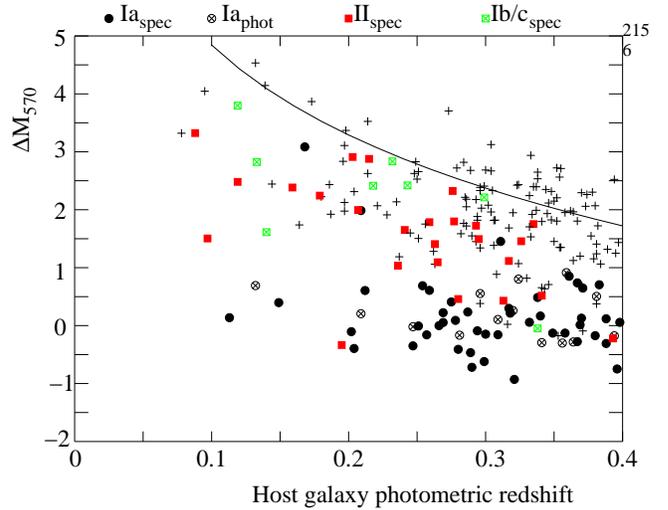}
\caption{ 
The pseudo-absolute magnitude
$\Delta M_{570}$ defined by (\ref{M600eq}) as a function of redshift.
Identified SNIa are concentrated near $\Delta M_{570}=0$.
The line in the upper-right corresponds to the
$m_{570}=24.1$.
Spectroscopically identified supernovae  and photometrically
identified SNIa are marked by the signs defined at the top of the figure.
}
\label{m600fig}
\end{figure}

\begin{figure}
\includegraphics[width=8.5cm]{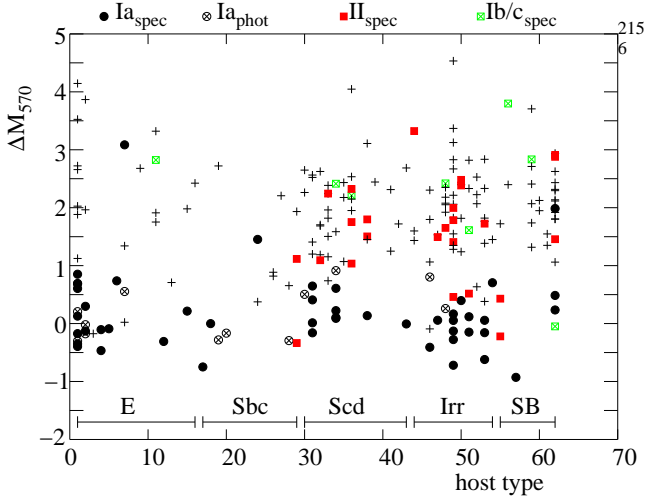}
\caption{Host morphological type as a function of  $\Delta M_{570}$.
The types are derived from the best spectral templates
of \citet{ilbert2006}.
Spectroscopically identified supernovae  and photometrically
identified SNIa are marked by the signs defined at the top of the figure.
}
\label{hosttypefig}
\end{figure}

\begin{figure}
\includegraphics[width=8.5cm]{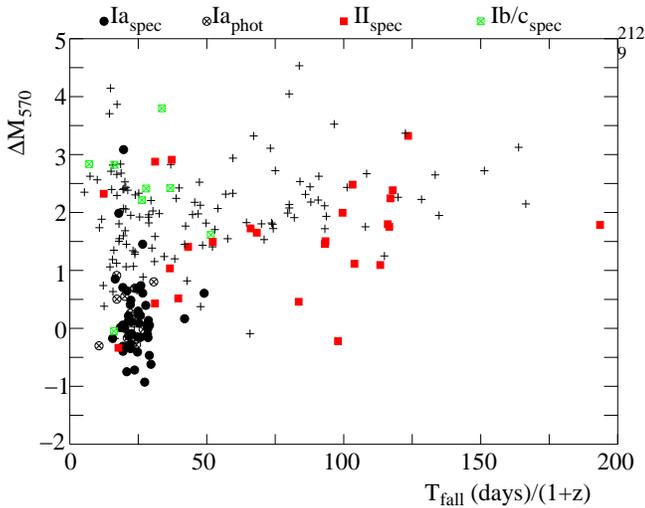}
\caption{ 
Fall time  $\tau_{fall}/(1+z)$  ($i^\prime$ band) as a function of  $\Delta M_{570}$.
Spectroscopically identified supernovae  and photometrically
identified SNIa are marked by the signs defined at the top of the figure.
}
\label{falltimefig}
\end{figure}

\begin{figure}
\includegraphics[width=8.5cm]{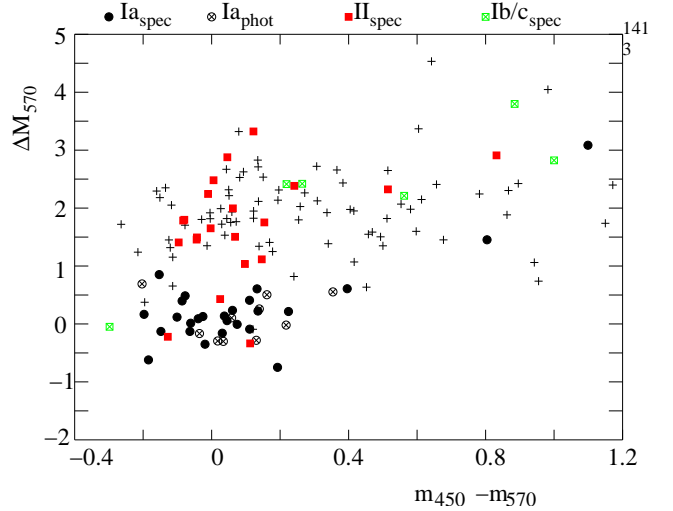}
\caption{ 
The color-magnitude diagram.  Only
the subset of 
events with points in $g^\prime$ near maximum light appear in the plot.
Spectroscopically identified supernovae  and photometrically
identified SNIa are marked by the signs defined at the top of the figure.
}
\label{colormagfig}
\end{figure}

Figure \ref{m600fig} shows $\Delta M_{570}$ as a function of redshift.
The spectroscopically identified SNIa and SNcc are now
separated horizontally with $\Delta M_{570}<0.75$ dominated by
SNIa and $\Delta M_{570}>0.75$ 
containing most spectroscopically-confirmed  SNcc.
The characteristics of the events as a function of $\Delta M_{570}$,
shown in Figures \ref{hosttypefig}-\ref{colormagfig},
are broadly consistent with those expected for SNIa and SNcc.
Figure \ref{hosttypefig} shows $\Delta M_{570}$ as a function
of the photometric host types \citep{ilbert2006}.
As expected for a sample dominated by SNcc,
the faint events have relatively fewer early-type hosts (19/152)
compared to 24/69 for the bright events.
Figure \ref{falltimefig} shows $\Delta M_{570}$
as a function of $\tau_{fall}/(1+z)$. 
As with low redshift SNcc \citep{richardson}, about half (47/108) the
faint events have  
$\tau_{fall}/(1+z)>50\,{\rm days}$, characteristic
of plateau SNII and significantly longer than fall times for
SNIa, $20<\tau_{fall}/(1+z)<30\,{\rm days}$.
Finally, Fig. \ref{colormagfig} shows the color-magnitude diagram
using 
the AB magnitude at $450\,{\rm nm}$ in the rest frame:
\begin{equation}
m_{450} \;\equiv\; (4z-0.4)r^\prime \,+\, (1.4-4z)g^\prime \;.
\end{equation}
The SNIa candidates have a narrower color distribution than the 
SNcc candidates.

\begin{figure}
\includegraphics[width=8.5cm]{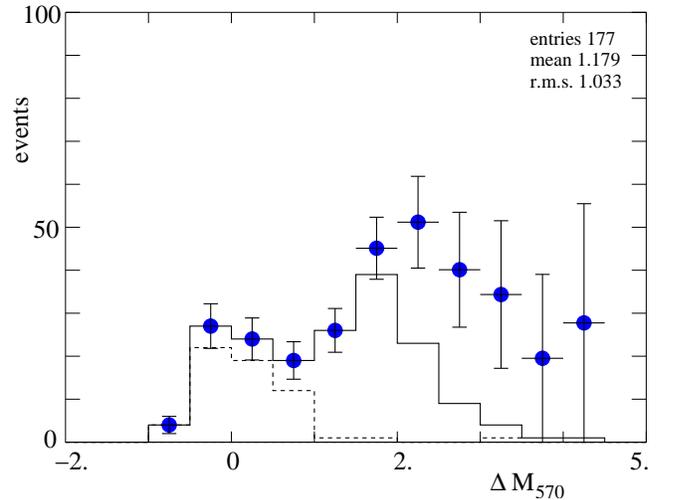}
\caption{The distribution of $\Delta M_{570}$.
The solid histogram  shows the distribution of $\Delta M_{570}$ for
events with $m_{570}<24.1$ and $0.05<z<0.4$.
The dashed histogram shows the distribution for spectroscopically
and photometrically selected
SNIa.
The data points with error bars 
show the number of events 
after correction for the detection efficiency and the 
dependence of the survey volume on $\Delta M_{570}$.
}
\label{ratefig}
\end{figure}

\begin{figure}
\includegraphics[width=8.5cm]{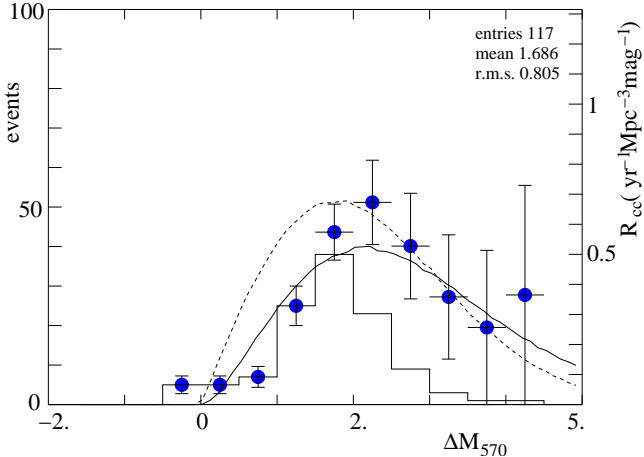}
\caption{
The differential SNcc rate.
The data points are those of Figure \ref{ratefig}
with the SNIa subtracted 
 (statistical errors only).
The left-hand scale is the number of events and
the right-hand scale is the absolute differential rate derived
from the total rate in Section \ref{ratesec}.
The dashed  curve shows the intrinsic distribution
for a toy model (of no physical motivation) with rate
$\propto \Delta M_{570}\exp[-( \Delta M_{570}/2.5)^2]$
while the solid line shows the distribution for
this model after including the host absorption
predicted by the model of
\citet{dustext}.
}
\label{absratefig}
\end{figure}

\section{Supernova classification}
\label{classificationsec}

\begin{table}[htb] \centering
  \begin{tabular}{ l l }
    \hline
                  & photometric  classification      \\
    spectroscopic  &                                   \\
    classification  & \hspace*{2mm} SNIa  \hspace*{5mm}   not SNIa     \\
    \hline
 &  \\
SNIa        & \hspace*{5mm}39   \hspace*{7mm} 7  \\
ambiguous   &\hspace*{5mm} 4    \hspace*{7mm} 24 \\
no spectrum & \hspace*{5mm}10   \hspace*{7mm} 69 \\
SNcc        & \hspace*{5mm} 1   \hspace*{7mm} 23 \\
 &  \\
\hline
  \end{tabular}
  \caption{
The numbers of events for each
spectroscopic and photometric classification.
}
  \label{classificationtab}
\end{table}

SNLS did not have sufficient telescope time to obtain spectra
of all SNIa candidates.
Therefore, in order to define a more complete SNIa sample,
the four-band light curves of all events were compared to 
SALT2 SNIa template light curves \citep{guysalt}.
The SALT2 model characterizes light curves by four parameters:
the date of maximum in the rest-frame B-band, the maximum flux in the
rest-frame B band, a ``color'' parameter roughly equivalent to rest-frame B-V,
and a ``stretch'' parameter that dilates the event time scale.
The light curves were fit for these parameters
imposing the host photometric redshift.
Events were ``photometrically'' classified as SNIa if
the four-band fit was reasonable ($\chi^2/dof<10$) and if 
fit parameters corresponded to normal SNIa.
In particular, cuts were applied to the rise and fall times, the 
color, $c$, and to the position in the two
color magnitude diagrams, 
$(g^\prime-i^\prime)\, vs.\, g^\prime$ and
$(r^\prime-z^\prime)\, vs.\, z^\prime$.

Table \ref{classificationtab} shows the 
number of events photometrically classified as SNIa or ``not SNIa''
for events classified spectroscopicaly as  SNIa, SNcc and
``ambiguous'', as well as for events for which no spectrum was obtained.
The table contains only those events
that will be used for rate measurements in the
next section, i.e. those with 
$0.05<z<0.4$ and $m_{570}<24.1$ (equation \ref{m570def}).
The reasonable performance of the  photometric classification  is
demonstrated by the fact that
only seven of the 46 spectroscopic SNIa were not selected 
and only one of the 24 spectroscopic SNcc was selected.
The photometric classification also selected 14 events that
had not been classified spectroscopically as SNIa.
As detailed below, the lack of spectroscopic confirmation
was generally due to an insufficient supernova signal over
the galactic background or to lack of telescope time to
obtain a spectrum.

As our nominal SNIa sample, we choose the 46 events that
were  spectroscopically identified as SNIa plus the 14 events
photometrically identified as SNIa that were not spectroscopic SNcc.
The nominal SNcc sample is the 117 remaining events.
The distribution of $\Delta M_{570}$ for these events are
shown as the histograms in Figs. \ref{ratefig} and \ref{absratefig}.
 
We have no evidence that the 60 SNIa candidates are contaminated
with SNcc incorrectly identified as SNIa.
We first consider the  46 spectroscopically identified SNIa.
It is unlikely that these events are significantly 
contaminated with SNcc since for $z<0.4$ the Si(615nm) line
is visible making the identification reliable.
In fact, of the 46 events, 43 were classified spectroscopically
as ``SNIa'' and only 3 as ``SNIa?''.  The three ``SNIa?'' events are all
selected photometrically making them good SNIa candidates.  
Of the 46 events, only seven were not photometrically accepted but
for reasons that do not call into question their SNIa character:
three had  photometric redshifts significantly different
from the spectroscopic redshifts causing the 
SALT fit to be very poor;  
one event had an extreme color parameter falling outside our
cuts; three events had a small number of poor  photometric points
causing the fits to fail our $\chi^2$ cut.

We now consider  contamination of the
14 SNIa that have only photometric confirmation.
Of these, four events have spectroscopy that was of
insufficient signal-to-noise  to determine
the SN type.
The remaining 10 events had no spectra either because
the event was discovered too late or because the estimated
signal-to-noise was insufficiently  (local flux increase $<20\%$).
Only one of the 14 events was judged ``unlikely'' to be a SNIa by
the spectroscopy target selection group, but the full
light curve indicates that it is consistent with being a
normal SNIa.
We therefore have no evidence that the 14 events are contaminated
with SNcc.
However, we have no good template catalog of bright SNcc to evaluate
the probability that a bright SNcc passes our SNIa photometric
selection.  We therefore
conservatively assign a systematic one standard deviation 
upper limit of $7\,{\rm events}$
to  contamination of the SNIa sample with SNcc.  

While we have no evidence that the SNIa sample is contaminated 
with SNcc, it is certain that the SNcc sample is contaminated
by sub-luminous SNIa.  
We will evaluate this contamination in the next section.

\section{The core-collapse rate}
\label{ratesec}

In this section, we will derive the SNcc rate using
events with $0.05<z<0.4$ and $m_{570}<24.1$.
The cut on  $m_{570}$ is used to ensure that only
events with good detection efficiency are used.
The requirement that $z>0.05$ eliminates one event at $z=0.04$. 
The uncertainty in  $\Delta M_{570}$ is   
$\sigma(\Delta M_{570})\sim 2\sigma_z/z$ so the
event with $z=0.04$ has $\sigma(\Delta M_{570})>1$ and we prefer to 
eliminate it.
(In fact, this event is a spectroscopic
outlier with $z_{spec}=0.247$.)
With the $m_{570}$ and redshift cuts, we are left with
177 events,
60 of which are spectroscopically
or photometrically identified SNIa.

From these numbers,
we will derive the SNcc rate as follows.
We first 
assign weights to the observed events 
that take into account detection
efficiency and the volume over which the event could
be detected by SNLS.
Because of their intrinsic faintness,
this will significantly increase the  number of  SNcc candidates  to 287.
We next evaluate two effects that can change the number of SNcc candidates
relative to SNIa candidates.
The first is simple 
spectral or photometric misidentification.
The second comes from the use of host photometric redshifts which,
we will see, has a slight tendency to increase the number of SNIa
candidates relative to SNcc candidates.
Using the corrected 
number of candidates, we then 
calculate  the SNcc rate relative to the SNIa rate.
By adopting the previously measured SNIa rate, we then
derive the SNcc rate for luminosities within $4.5\,{\rm mag}$
of normal SNIa.
Finally, we estimate the total SNcc rate taking into account 
the decrease in the number of observed supernovae due to extinction by dust
in the host galaxy.

\subsection{Event weights}

The observed distribution of $\Delta M_{570}$ is 
the histogram shown in Fig. \ref{ratefig}
for the 177 events with $z<0.4$ and $m_{570}<24.1$.  
In order to derive the true distribution of $\Delta M_{570}$
for events with $z<0.4$,
this distribution must be corrected for the $i^\prime$-dependent
detection efficiency and, more importantly, for the fact that
an event with absolute
magnitude $\Delta M_{570}$ can be seen only up to a 
redshift, $z_{max}$ defined by 
\begin{equation}
\sqrt{1+z_{max}}d(z_{max}) \;=\;10^{0.2(m_{570max}-C-\Delta M_{570})}
\end{equation}
where $d(z)$ is defined by (\ref{ddefeq})
and $m_{570max}=24.1$ is the maximum accepted magnitude.
Events with $z_{max}<0.4$ (i.e. $\Delta M_{570}>1.7$)
must be given 
a weight, $W>1$.  
The weight takes into account, most importantly,
the fact that SNLS detects them over a smaller volume.
Of secondary but non negligible importance is the fact
that the SNcc rate is believed to increase with redshift,
$R_{cc}\propto (1+z)^\alpha$ with $\alpha\sim3.6$ to reflect
the increasing star-formation rate with redshift.
We are therefore sensitive to intrinsically faint supernovae 
only in a redshift range where rate is small.
We correct for this with $\alpha=3.6$ appropriate for SNcc because
our SNIa candidates are all bright enough to have $z_{max}>0.4$.
Finally, because of cosmological time dilation, the
SNLS observing time is proportional to $(1+z)^{-1}$.
The total weight, $W(\Delta M_{570},i^\prime)$, is therefore given by
\begin{equation}
W^{-1} 
\;\propto\;\frac{\epsilon(i^\prime)}{\epsilon(i^\prime=21)}
 \int_0^{z_{max}}
\frac{dz^\prime d(z^\prime)^2 (1+z^\prime)^{\alpha-1} }
{\sqrt{\Omega_m (1+z^\prime )^3 + (1-\Omega_m)}} \;,
\label{weightdefeq}
\end{equation}
where $\epsilon(i^\prime)$ is the event detection efficiency.
The factor of proportionality is chosen so that 
$W(z_{max}=0.4,i^\prime=21)=1$.
Without the factors of $(1+z)$ in (\ref{ddefeq}) and 
(\ref{weightdefeq}), the weight would  be simply the product of 
$\epsilon(i^\prime=21)/\epsilon(i^\prime)$ and  the euclidean volume
ratio $(0.4/z_{max})^3$.
The factors of $(1+z)$ correct for the redshift
evolution of the volume element, the exposure time, and the SNcc rate.

Weighting individual events  gives 
the corrected $\Delta M_{570}$ distribution
shown by the  data points and error bars in Fig. \ref{ratefig}.
All of the 60 SNIa candidates have  weights near unity.
Because of their faintness, many of the 117 SNcc candidates have
$W>1$ and the corrected number of SNcc candidates 
is 
$287\pm40$  (statistical error only).

\subsection{Corrections for misidentification and redshift migration}

In this section we correct the raw number of SNIa and SNcc candidates
for two effects that can affect their numbers:  type misidentification
(summarized in Table \ref{correctiontab})
and redshift migration 
due to the use of photometric redshifts.

\begin{table} \centering
  \begin{tabular}{ l l l }
    \hline
 &  & \\
   correction  &  SNIa & SNcc   \\
 &  & \\
    \hline
 &  & \\
 SNIa incorrectly identified as SNcc & & \\
\hspace*{5mm}sub-luminous SNIa       & $+9\pm5$  & $-9\pm5$ \\
\hspace*{5mm}normal SNIa & $+2\pm1$ & $-2\pm1$ \\
 & &  \\
SNcc incorrectly identified as SNIa    & $0^{+0}_{-7}$ & $0^{+7}_{-0}$\\
 &  & \\
 Contamination by non-SN & $0\pm0$ & $0^{+0}_{-4}$ \\
     &  & \\
 Total & $+11^{+5}_{-9}$ &  $-11^{+9}_{-7}$ \\
 &  & \\
\hline
  \end{tabular}
  \caption{
Corrections applied to the 60 SNIa candidates
and 287 (weighted) SNcc candidates. 
}
  \label{correctiontab}
\end{table}

The first shift in the SNIa-SNcc ratio
is due to SNIa that are incorrectly
identified as SNcc.
We divided this correction into that for ``sub-luminous'' SNIa and
normal SNIa.
Sub-luminous SNIa \citep{sublum}  have a mean
magnitude 1.5mag below the mean magnitude for normal SNIa and
account for $16\pm6\%$ of  SNIa.
None are found in our sample since both selection for spectroscopy
and photometric selection aimed at finding normal SNIa.
We therefore add (subtract) $9\pm5\,{\rm events}$ to the SNIa (from
the SNcc) samples.
For normal SNIa, we must
correct for events that were neither spectroscopically
nor photometrically selected.
From Table \ref{classificationtab},
of the 46 spectroscopically confirmed events, only 7 were not
photometrically selected.
This  gives an inefficiency of $7/46=0.15$
for photometric identification of spectroscopically confirmed SNIa.
To the 14 SNIa candidates relying solely on photometric
selection, we can therefore add $0.15\times14=2\,{\rm events}$ and
subtract the same number from the SNcc.

As discussed in Section \ref{classificationsec}, we make no
correction for SNcc incorrectly identified as SNIa but 
assign a systematic one standard deviation 
upper limit of $7\,{\rm events}$
to  contamination of the SNIa sample with SNcc.  

Contamination with non-supernova events 
is expected to be unimportant.  
The scan of events resulted in the elimination of only six AGN-like
events
and the identification of four additional
events that were judged uncertain.
We adopt four events as our one standard deviation upper limit
on AGN contamination of the SNcc sample.

Finally, we correct for redshift migration (Eddington bias), an effect that
comes from our use of photometric redshifts with a modest
resolution of $\sigma_z\sim 0.04$.
Because there are more supernovae at high redshift than
at low redshift, the main effect of this resolution is
for high redshift supernovae to migrate below the $z=0.4$ cutoff.
If there were no cut  $m_{570}<24.1$,
this would increase the number of SNIa and SNcc
candidates by the same factor.
The fact that SNcc are fainter than SNIa means that migrating
SNcc are less likely to satisfy  $m_{570}<24.1$ than
migrating SNIa.
We have used a Monte Carlo simulation to estimate this effect.
The simulation generates events with a realistic redshift
and $M_{570}$ distribution and uses the observed spectroscopic-photometric
redshift pairs from Fig. \ref{zvszfig} to assign photometric
redshifts.  Outliers in this plot are used so the simulation
takes into account catastrophic redshifts.
Counting weighted simulated events indicates that the
migration makes the measured SNcc-SNIa rate ratio $(15\pm4)\%$
less than the real rate ratio.
The statistical error comes from the limited number of redshift
pairs we have used for the simulation.
The measured SNcc-SNIa rate will therefore be multiplied by a factor
1.15 to take into account this effect.

\subsection{The SNcc-SNIa relative rate}

The corrections for the number of events shown in Table \ref{correctiontab}
give an increase of  $11^{+5}_{-9}$ SNIa candidates,
and a corresponding decrease in the number of SNcc candidates.
The corrected relative rate for $z<0.4$ is therefore
\begin{displaymath}
\frac{R_{cc}(\Delta M_{570}<4.5)}{R_{Ia}}
 =
\frac{287-11}{60+11} \times 1.15
\end{displaymath}
\begin{displaymath}\hspace*{20mm}
\,=\, 4.5\pm0.8(stat.)\, ^{+0.9}_{-0.7} (sys.) \;.
\end{displaymath}
where the factor 1.15 takes into account redshift migration.
The ratio is for $z<0.4$ corresponding to an expected  mean redshift
of $0.306$ for a rate proportional to $(1+z)^2$ and a mean
of $0.313$ for a rate proportional to $(1+z)^{3.6}$.
Our sample of 60 SNIa has
a mean redshift of $0.30\pm0.01$, consistent with expectations
for a complete (volume limited) sample.

The systematic error in $R_{cc}/R_{Ia}$
includes those due to the corrections
from the previous section as well as three 
additional systematic
uncertainties which we add in quadrature.

The first additional systematic
concerns  the uncertainty in the relative efficiencies for SNIa and
the fainter SNcc.
To avoid large uncertainties, we have used only events with $m_{570}<24.1$
where the efficiency is high. 
With this cut, there is only a 10\% difference in the SNcc rate
calculated with the nominal efficiencies and that 
calculated assuming a magnitude-independent efficiency.
We adopt 10\% as the nominal systematic error from this source.
To check that
there is no significant uncorrected event loss near the magnitude cut,
we verified that the derived SNcc rate does not depend significantly
on the position of the magnitude cut.
For example, using  $m_{570max}=23.6$, the number of events
SNcc candidates is reduced from 
117 to 82.
After weighting, this is increased to
$334\pm80$
consistent with the
 $287\pm45$ event found using $m_{570max}=24.1$.
(Most of the increase comes from the two events with
$\Delta M_{570}>3.5$ which are given greater weights
with $m_{570max}=23.6$).

The second systematic concerns the star-formation rate.
The corrected differential rate was calculated assuming
that the SNcc rate is proportional to $(1+z)^\alpha$ with
$\alpha=3.6$ according to \citet{hopkins}.
These authors do not cite an uncertainty for $\alpha$ but inspection
of the data indicates that $\alpha=3.6\pm1.0$ is reasonable.
This corresponds to a 10\% systematic uncertainty in the SNcc rate.

The final systematic concerns our requirement
that a host galaxy be found and that a redshift
be given in the \citet{ilbert2006} catalog.
This requirement could conceivably favor SNcc or SNIa.
For spectroscopically identified supernovae with $z<0.4$,
a host galaxy is generally found but a redshift may not
be given in the catalog.
For 41 spectroscopic SNIa
with spectroscopic redshifts $<0.4$ that
were found in the deferred search,
only 4 have no host redshift
while for the 36 spectroscopic SNcc there are only 3 with
no host redshift.
Thus, we see no difference in host-redshift measurement
efficiency at the 5\% level and we adopt this as 
the systematic uncertainty.

\subsection{The SNcc rate}

To derive a value of $R_{cc}$ we adopt the value of $R_{Ia}$
measured by \citet{neill2006} at $z=0.5$:
$0.42\times 10^{-4} \yr^{-1}(h_{70}^{-1}\Mpc)^{-3}$
Our measurement of $R_{cc}/R_{Ia}$ is effectively
at $z=0.3$ 
and we adopt a SNIa rate at this redshift of
$0.315\times 10^{-4} \yr^{-1}(h_{70}^{-1}\Mpc)^{-3}$
calculated assuming  $R_{Ia}\propto(1+z)^2$.
This gives a SNcc rate
within 4.5 magnitudes of normal SNIa of 
\begin{displaymath}
\frac{R_{cc}(\Delta M_{570}<4.5)}
{10^{-4}\yr^{-1}(h_{70}^{-1}\Mpc)^{-3} }  \;= 
1.42\pm 0.30\,(stat.)\,^{+0.32}_{-0.24}\,(sys.)
\end{displaymath}
We have added the statistical and systematic uncertainties
of  $R_{cc}/R_{Ia}$ and of $R_{Ia}$ separately in quadrature
though not including the systematic uncertainty in $R_{Ia}$ due
to sub-luminous and absorbed supernovae because it is already
included in the uncertainty in  $R_{cc}/R_{Ia}$.

With the determination of the total SNcc rate, we can give
an absolute differential rate per unit magnitude for SNcc.
It is shown as the right-hand scale in Fig. \ref{absratefig}.
The rate is measured down to luminosities $4.5\,{\rm mag}$
fainter than normal SNIa.  
It should however be emphasized that
there are only two events with $\Delta M_{570}>3.5$.
One of them has spectroscopic confirmation
and the spectroscopic redshift, $z=0.131$, is in good
agreement with the host photometric redshift, $z=0.119$.
The other event has a host 
spectroscopic redshift\footnote{http://nedwww.ipac.caltech.edu/index.html}
$z=0.0815$
in good agreement with the host photometric redshift used
here, $z=0.095$.
Thus, we have no indication that these two events are higher
luminosity events that have migrated from high redshift.

To estimate
a {\it total} rate for SNcc we need to estimate
the number of SNcc with  $\Delta M_{570}>4.5$
either because they are intrinsically faint 
(e.g. SN1987A, $\Delta M_{570}\sim 5.5$) or
because of high host extinction.
SNLS obviously cannot say anything about intrinsically faint supernovae.
However,
by adopting a host galaxy extinction model, we can
estimate the number of SNcc that have intrinsic
luminosities within our range of sensitivity but that are
lost because of high host extinction.
We have used the results of
\citet{dustext} who give (their table 1 and Figure 1)
the distribution of $A_B$
as a function of host inclination angle.
This can be converted to a distribution of absorption
at $570\,{\rm nm}$ and convoluted with the
pre-absorption distribution of $M_{570}$.
For example, if we model the intrinsic SNcc
magnitude distribution 
shown as the dashed line in Fig. \ref{absratefig},
then
the SNcc host extinction  model of \citet{dustext} predicts 
the distribution shown by the solid line in the Figure.
With this model,
15\% of SNcc have $\Delta M_{570}>4.5$.
Our estimated total rate is then increased to 
$R_{cc}=1.63 \times10^{-4}\yr^{-1}(h_{70}^{-1}\Mpc)^{-3}$.
In our model, most of the events with $\Delta M_{570}>4.5$
are highly absorbed so our estimate 
should be considered a lower limit on the SNcc rate that
ignores supernovae that
are intrinsically fainter than  $\Delta M_{570}=4.5$.

\section{Discussion}
\label{sumsec}

\begin{figure}
\includegraphics[width=8.5cm]{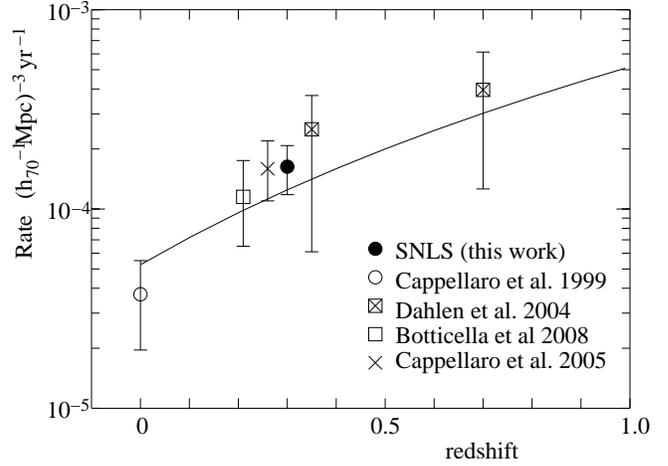}
\caption{
The measured rate of SNcc as a function of redshift.
The SNLS point includes a 15\% correction for host absorption
as described in the text.
The error bars correspond to statistical and systematic
uncertainties added in quadrature.
The line is the best fit for rate$\propto(1+z)^{3.6}$, i.e.
proportional to the SFR.
}
\label{ratesumfig}
\end{figure}

Figure \ref{ratesumfig} summarizes the published measurement
of the SNcc rate.
All data is consistent with a rate that increases with redshift
like the  SFR  $\propto(1+z)^{3.6}$.
It should be emphasized that the previous measurements
use quite different detection and analysis procedures.
We therefore refrain from drawing any quantitative 
conclusions about
the redshift dependence of the SNcc rate.

Our results will be improved in the future with the addition 
of two more years
of SNLS data, and with the use of host spectroscopic redshifts that we are
in the process of obtaining.

\begin{acknowledgements}

The SNLS collaboration gratefully acknowledges the assistance of Pierre
Martin and the CFHT Queued Service Observations team.
Canadian collaboration members acknowledge support from NSERC and
CIAR; French collaboration members from CNRS/IN2P3, CNRS/INSU and CEA.

SNLS relies on observations with MegaCam, a joint project of
CFHT and CEA/DAPNIA, at the Canada-France-Hawaii Telescope (CFHT)
which is operated by the National Research Council (NRC) of Canada, the
Institut National des Science de l'Univers of the Centre National de la
Recherche Scientifique (CNRS) of France, and the University of Hawaii. This
work is based in part on data products produced at the Canadian
Astronomy Data Centre as part of the Canada-France-Hawaii Telescope Legacy
Survey, a collaborative project of the National Research Council of
Canada and the French Centre national de la recherche scientifique.
\end{acknowledgements}

\bibliographystyle{aa}

\end{document}